# Empirical 3D Channel Modeling for Cellular-Connected UAVs: A Triple-Layer Machine Learning Approach

HAIDER A.H. ALOBAIDY[1], MEMBER, IEEE, MEHRAN BEHJATI[2], SENIOR MEMBER, IEEE, ROSDIADEE NORDIN[2], SENIOR MEMBER, IEEE, MUHAMMAD AIDIEL ZULKIFLEY[3], NOR FADZILAH ABDULLAH[4], MEMBER, IEEE, AND NUR FAHIMAH MAT SALLEH[3]

[1] Department of Information and Communications Engineering, College of Information Engineering, Al-Nahrain University, Baghdad, Iraq
[2] Faculty of Engineering & Technology, Sunway University, No. 5, Jalan Universiti, Bandar Sunway 47500, Selangor, Malaysia
[3] Argentavis, Aerodyne Group, Persiaran Cyber Point Selatan, Cyber 8, 63000 Cyberjaya, Selangor, Malaysia
[4] Department of Electrical, Electronic & Systems Engineering, Universiti Kebangsaan Malaysia, Bangi 43600, Selangor, Malaysia

CORRESPONDING AUTHOR: Mehran Behjati (e-mail: mehranb@sunway.edu.my).

This work was supported by Collaborative Research in Engineering, Science and Technology (CREST) and Aerodyne Group under grant reference number T23C2-19. We also acknowledged various forms of contributions from Universiti Kebangsaan Malaysia and Sunway University in getting this research work being published.

**ABSTRACT** This work proposes an empirical air-to-ground (A2G) propagation model specifically designed for cellular-connected unmanned aerial vehicles (UAVs). An in-depth aerial drive test was carried out within an operating Long-Term Evolution (LTE) network, gathering thorough measurements of key network parameters. Rigid preprocessing and statistical analysis of these data produced a strong foundation for training a new triple-layer machine learning (ML) model. The proposed ML framework employs a systematic hierarchical approach. Accordingly, the first two layers, Stepwise Linear Regression (STW) and Ensemble of Bagged Trees (EBT) generate predictions independently; meanwhile, the third layer, Gaussian Process Regression (GPR), explicitly acts as an aggregation layer, refining these predictions to accurately estimate Key Performance Indicators (KPIs) such as Reference Signal Received Power (RSRP), Reference Signal Received Quality (RSRQ), Received Signal Strength (RSSI), and Path Loss (PL). Compared to traditional single-layer ML or computationally intensive ray-tracing approaches, the proposed triple-layer ML framework significantly improves predictive accuracy and robustness, achieving around 99% accuracy in training and above 90% in testing while utilizing a minimal but effective feature set—log-transformed 3D and 2D propagation distances, azimuth, and elevation angles. This streamlined feature selection substantially reduces computing complexity, thus enhancing scalability across various operating environments. The proposed framework's practicality and efficacy for real-world deployment in UAV-integrated cellular networks are further demonstrated by comparative analyses, which underscore its substantial improvement.

**INDEX TERMS** Cellular-connected drones, UAV, aerial communications, wireless communications, channel modeling, machine learning, propagation Characteristics, path loss model

## I. INTRODUCTION

The rapid proliferation of unmanned aerial vehicles (UAVs) has driven significant interest in integrating them into existing cellular communication infrastructures, supporting applications such as payload delivery, surveillance, emergency response, and real-time multimedia streaming [1], [2]. Reliable communication is essential for UAVs to operate safely and efficiently, especially in scenarios requiring real-time interaction with ground or aerial infrastructure [2], [3]. However, UAVs exhibit unique operational characteristics, including high-altitude mobility, dynamic flight trajectories, and rapidly varying propagation environments, that introduce complex propagation challenges not effectively addressed by traditional terrestrial models [3], [4]. These distinct conditions necessitate robust and efficient propagation models that accurately reflect the aerial cellular channel characteristics using flexible and adaptive methods, such as machine learning (ML).

UAV communication systems rely predominantly on satellite communications, dedicated communication links, radio-frequency (RF)-based systems, and cellular networks. Satellite and dedicated systems provide wide-area coverage and reliable communications but suffer from high latency, limited scalability, high operational cost, and infrastructure complexities [5], [6]. While cost-effective and easy to deploy, RF-based systems face inherent limitations like interference, limited operational range, and multipath-induced performance degradation [3]. In contrast, cellular networks leveraging widespread deployment of LTE and emerging 5G infrastructures provide a promising communication solution, offering ubiquitous coverage, scalability, and higher data rates suitable for UAV operations.

With inherent features such as low latency, high reliability, and massive machine-type communications (mMTC), cellular networks are increasingly considered the most practical and cost-effective option for supporting diverse UAV applications [7]. Despite these advantages, integrating UAVs into terrestrial-focused cellular networks presents substantial technical challenges, particularly in accurately modeling air-to-ground (A2G) propagation characteristics [8].

UAV channels exhibit dynamic variations due to continuous changes in altitude, elevation and azimuth angles, propagation distance, and unpredictable line-of-sight (LoS) or non-line-of-sight (NLoS) conditions [3], [4]. Traditional empirical propagation models often oversimplify these dynamic aspects, employing static assumptions that fail to capture the realistic variability encountered in aerial scenarios. Similarly, although accurate in controlled environments, ray-tracing or geometry-based modeling approaches are computationally intensive and heavily reliant on precise environmental data, making them impractical for scalable real-world deployments [9], [10].

Recent research has explored ML methods to overcome limitations inherent in traditional propagation modeling. Leveraging empirical measurements and minimal environmental data, ML techniques offer promising solutions to dynamically predict critical network parameters like Reference Signal Received Power (RSRP), Reference Signal Received Quality (RSRQ), Received Signal Strength (RSSI), and Path Loss (PL). However, current ML-driven propagation models often utilize single-layer architectures, limiting their ability to refine predictions systematically. Moreover, existing ML models frequently depend on overly complex or extensive feature sets, negatively affecting their computational efficiency and scalability for real-time UAV communications [10], [11]. Therefore, there remains an essential need for a predictive propagation modeling framework that balances predictive accuracy, minimal feature complexity, and scalability across diverse UAV operating conditions.

This study introduces a novel triple-layer ML propagation modeling framework for cellular-connected UAV communications to address these critical research gaps. The proposed framework employs a systematic prediction refinement approach, integrating three distinct ML methods: a stepwise linear regression model (STW), an ensemble of bagged trees (EBT), and Gaussian process regression (GPR) as an aggregation layer. Unlike single-layer ML models, this multi-layer architecture significantly improves prediction accuracy and robustness against dynamic variations in the UAV's operational parameters. Moreover, the proposed model utilizes a minimal yet effective input feature set, specifically, log-scale 3D and 2D distances, elevation, and azimuth angles.

The key contributions of this study are as follows:
1. A triple-layer ML framework is introduced, consisting of STW model for initial regression, an EBT for refinement, and a final aggregation layer to fine-tune predictions. By adding the triple-layer approach, this framework achieves a prediction accuracy of approximately 97%.
2. A minimal yet effective set of input features (log-transformed 3D and 2D distances, azimuth, and elevation angles) is employed, significantly simplifying computational complexity while maintaining high accuracy compared to traditional ML models.
3. A comprehensive dataset of 11,060 aerial communication samples, filtered from an original set of 95,000 samples, is publicly available at [12]. The dataset's diversity in flight scenarios and network conditions ensures the generalizability of the proposed models.
4. The proposed ML framework offers a practical balance between computational efficiency, scalability, and predictive accuracy, making it highly suitable for real-world deployment within cellular-connected UAVs.

The remainder of the paper is organized as follows: Section II reviews advancements and highlights specific limitations of existing ML-based propagation models. Section III details the measurement setup and data collection methodology. Section IV describes the proposed triple-layer ML modeling framework. Section V presents detailed results and discussions, and Section VI concludes the paper and identifies avenues for future research.

## II. ADVANCES IN CHANNEL MODELING AND RELATED WORKS

Accurate and efficient channel modeling ensures reliable UAV communication within cellular networks. In contrast to terrestrial communication, UAVs experience distinct propagation challenges due to their varying altitudes, high mobility, and dynamic LoS conditions [13]–[17]. These unique characteristics pose considerable challenges for traditional empirical propagation models primarily developed from terrestrial and relatively static communication scenarios. Consequently, these empirical models struggle to generalize effectively to UAV-specific environments. Deterministic modeling methods, such as ray tracing, explicitly simulate propagation paths and offer improved accuracy yet require detailed environmental information and intensive computations, limiting their practical scalability in real-world UAV deployments.

Meanwhile, although introduced to balance accuracy and complexity, hybrid approaches combining deterministic and empirical or ML-based methods may still yield computational complexities that limit scalability. To overcome these limitations, there is a pressing need to adopt ML techniques, leveraging empirical measurements and minimal environmental parameters to model UAV propagation effectively. This section provides a focused review of key advancements and identifies critical gaps in empirical measurements, ML-driven modeling, and available datasets, motivating the triple-layer ML approach proposed in this work.

### A. EMPIRICAL MEASUREMENTS AND DATASETS

Empirical datasets are the cornerstone for developing and validating ML-based channel models, providing real-world insights into UAV propagation behavior. Several measurement campaigns have been conducted to characterize A2G channel conditions in cellular networks. For example, [18] analyzed UAV downlink signals from an operational LTE network at 2.585 GHz, recording channel impulse responses (CIRs) across different flight paths. The study employed the Space-Alternating Generalized Expectation-Maximization (SAGE) algorithm to extract multipath component parameters, generating stochastic models for path loss, shadow fading, and delay spread. However, this work was limited to specific suburban environments and did not explore generalized ML-based predictions.

The Addis dataset in [9] introduced a volumetric radio environment mapping (VREM) approach, offering 3D spatial propagation loss maps at various altitudes. Although valuable for static environment modeling, it lacks dynamic flight data and real-world UAV mobility patterns, restricting its applicability to practical UAV communication scenarios. Similarly, Maeng et al. [19] developed the LTE I/Q dataset using UAV flights at 30–110 m altitudes via the NSF AERPAW platform, offering raw I/Q samples for ML-driven channel modeling. Despite addressing some limitations of static datasets, this dataset does not account for key parameters such as azimuth and elevation angles, which are essential for accurate ML-based 3D propagation modeling.

Recent empirical studies, such as [20], employed UAV drive tests to analyze radio link parameters in commercial 5G networks. These tests provide real-world insights into UAV cellular connectivity; however, they remain focused on signal strength characterization rather than predictive modeling using ML. Collectively, existing datasets and empirical studies highlight the need for an extensive, diverse dataset specifically designed for ML-based propagation modeling, incorporating key parameters such as 3D distances, azimuth, and elevation angles to improve UAV channel prediction accuracy.

### B. ML-BASED APPROACHES IN UAV CHANNEL MODELING
Machine learning techniques have emerged as powerful tools for UAV channel modeling, leveraging empirical datasets to enhance prediction accuracy. As in [21], unsupervised learning and clustering methods have been employed to classify UAV signal quality for A2G communication scenarios based on RSSI. The method was then utilized to construct rapid 3D temporary channel models, achieving up to 91.8% similarity with traditional statistical models. Although promising, heavy reliance on clustering approaches may result in a lack of predictive capabilities for real-time UAV channel estimation and often a struggle with scalability across diverse flight conditions.

More advanced ML applications have been explored in [22] for UAV-to-Vehicle (U2V) millimeter-wave (mmWave) channel modeling, where backpropagation neural networks (BPNN) and generative adversarial networks (GANs) have been utilized alongside extensive ray-tracing simulations. While these models provide high prediction accuracy, their reliance on synthetic data and complex feature extraction limits real-world generalization. Additionally, their sensitivity to UAV orientation changes remains a significant limitation, as roll and pitch angles substantially impact signal propagation in practical deployments.

Bahjati et al. [17] developed six ML-based models for UAV-based key performance indicators (KPI) estimation to predict RSRP and RSRQ within LTE networks. These models successfully demonstrated ML's potential in aerial coverage assessment and reliability, a crucial factor for missions beyond visual line of sight (BVLOS). Although the proposed models deliver strong performance, the reliance on single-layer ML models' architecture limits their prediction refinement and optimization applicability to more intricate environments. Similarly, Bithas et al. [23] surveyed ML-driven techniques for UAV communication, covering areas such as channel modeling, resource management, positioning, and security, yet lacked a dedicated multi-layer ML approach for comprehensive A2G channel modeling.

ML has also been applied to specific tasks such as direction estimation for air-to-air (A2A) links, as in [24], where up to 86% accuracy was achieved in beam selection. However, these models primarily focus on beamforming rather than generalized channel modeling for UAVs. Moreover, emerging research on terahertz (THz) bands and reconfigurable intelligent surfaces (RIS) for UAVs has demonstrated potential for high-data-rate aerial communications [25], [26]. However, these methods require precise environmental control and are highly sensitive to UAV orientation, limiting their applicability in diverse urban and suburban settings.

### C. DISCUSSION AND RESEARCH GAPS
Despite the advancements in empirical and ML-based channel modeling techniques for cellular-connected UAVs, several critical limitations remain unresolved in existing literature. Firstly, current ML-driven propagation models predominantly adopt single-layer architectures, limiting their capacity for systematic prediction refinement and reducing accuracy in dynamic, real-world aerial scenarios. Secondly, existing models often rely on extensive feature sets or complex data preprocessing, leading to increased computational complexity that hampers practical scalability. Lastly, publicly available empirical datasets typically lack diversity in UAV flight scenarios and critical propagation parameters (e.g., azimuth, elevation angles), significantly restricting model validation, generalizability, and reproducibility.

Therefore, a compelling need exists to explore a more systematic and simplified ML-driven approach that directly addresses these gaps. In this context, a hierarchical, multi-layer ML framework emerges as a promising direction. Such a model would allow progressive refinement of predictions by initially leveraging simpler predictive layers and subsequently combining their outputs through an aggregation layer. Crucially, this method could reduce computational complexity by employing a minimal yet highly informative set of features, such as log-transformed propagation distances and angular parameters.

Motivated by these observations, this study investigates and proposes a triple-layer ML framework specifically designed to overcome the identified limitations. The detailed architecture, technical implementation, and validation procedures of this approach will be thoroughly introduced and discussed in the next sections.

### III. MEASUREMENT SETUP AND DATA COLLECTION
To develop accurate prediction models that reflect real-world cellular radio propagation characteristics, our methodology began with conducting aerial drive tests. Subsequently, these measurements were utilized to assess the performance of the cellular link in a UAV-to-BS communication scenario. This section outlines the specifics of the measurement setup and methods employed, details the study area, and elucidates the strategies utilized to validate the obtained results.

## A. DRONE FLIGHT MISSION PREPARATIONS

Fig. 1 illustrates the overall measurement approach, utilizing the DJI Matrice 300 (M300) drone for the aerial drive test. The study required a smartphone with a specific drive-test application to measure cellular KPIs. For this purpose, the Keysight Nemo Handy [27] software was selected mainly due to its superior accuracy and comprehensive capability in monitoring essential KPIs within the cellular network, surpassing other available applications [8]. The application was preinstalled on a Samsung Galaxy S21+ phone, functioning as standard user equipment (UE) within the network. Keysight Nemo provides specialized software offering extended scripting and logging functionalities, facilitating direct recording of KPIs from the chipset. Consequently, the Keysight software accurately assesses and portrays the quality of service encountered by a genuine UE.

The smartphone was mounted on the drone and served by the LTE network during the measurement. At the same time, Keysight Nemo measured LTE-related parameters such as drone (or UE) position, RSRP, RSRQ, RSSI, and PL. The measured KPIs were then stored on the phone and retrieved after landing. The carrier box was specifically designed to maintain drone stability and operational safety, and it was made of cardboard to avoid signal attenuation.

To achieve high-resolution KPI sampling, the drone maintained a constant speed of 20 km/h throughout the tests. Given that Keysight Nemo samples at a rate of 2 Hz, the UE recorded data approximately every 5 meters. Additionally, this speed helped minimize Doppler effects and ensure consistent signal measurements.

All aerial operations have obtained airspace regulatory approval and adhered to regulations set by the Civil Aviation Authority of Malaysia (CAAM) [28]. Compliance with CAAM safety guidelines was strictly observed, including maintaining Visual Line of Sight (VLOS), adhering to altitude restrictions, monitoring battery levels, and performing pre-flight safety checks.

The flight path was carefully designed to ensure smooth navigation, consistent flight dynamics, and accurate data sampling. After completing each predefined waypoint route, the drone was configured to hover briefly before proceeding with subsequent missions, ensuring stable transitions and accurate positioning data logging. Flight altitudes were precisely referenced to Above Sea Level (ASL) and Above Ground Level (AGL), starting at 85 m ASL (10 m AGL) and incrementally increasing by 5 m up to 205 m ASL (130 m AGL). At each altitude, measurements were taken at radial distances ranging from 100 m to 900 m in 100 m increments.

To optimize data quality and operational efficiency, flight missions were carefully structured to maximize drone endurance while ensuring consistent and high-resolution data collection. Each battery pack provided approximately 40 minutes of effective flight time, allowing extended measurement sessions before requiring a swap. The data collection campaign was conducted over five operational days, accumulating a total flight time of approximately 14 hours. This structured approach ensured comprehensive spatial and temporal coverage, facilitating robust analysis of the air-to-ground propagation characteristics.

## B. DRIVE TEST SETUP

The measurement was conducted at the National University of Malaysia (UKM), a tropical suburban metropolitan area characterized by undulating terrain and dense vegetation. Fig. 2 presents an aerial image of the study site.

In the considered study area, cellular coverage is provided by different telecommunication service providers, and in this study, only one of the cellular operators with better coverage was selected for the measurement. The considered LTE BS was equipped with four LTE sectors with a carrier frequency of 2.6 GHz. Each antenna sector featured horizontal and vertical 3 dB beamwidths of approximately 105° and 6°, respectively. Fig. 3 depicts areas under the coverage of different sectors.

Drive tests were conducted at different heights from 10–130 m AGL, with increments of 5 m, and at radial distances of 100–900 m with 100 m steps. The drone followed a predefined partially circular path, as shown in Fig. 4. The measurement scenario depicted in Fig. 1 and Fig. 4 clearly illustrates the systematic curved flight paths at incremental altitudes relative to the BS antenna height, which is situated at approximately 85 m ASL (10 m AGL).

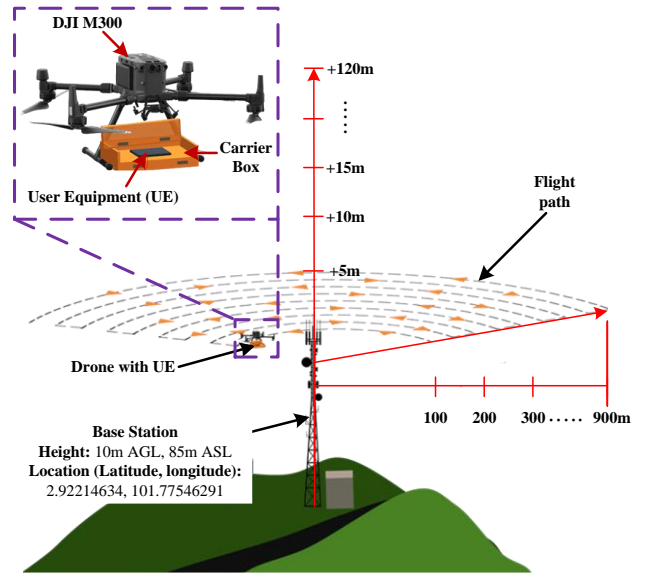

**FIGURE 1.** Illustration of the aerial drive test method and depiction of utilized drone and the cardboard box carrying the UE.

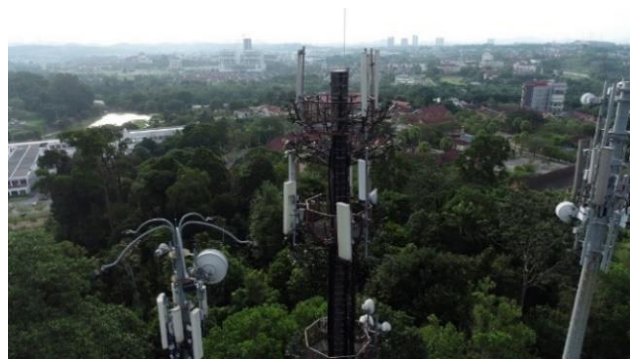

**FIGURE 2.** Overview of the suburban study environment for the aerial drive test.

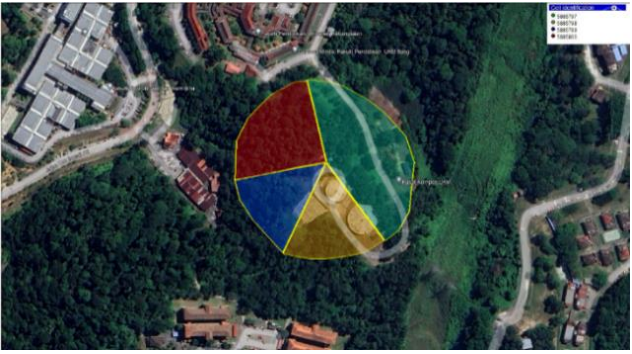

FIGURE 3. Estimated coverage of each of the four sectors at the considered LTE base station.

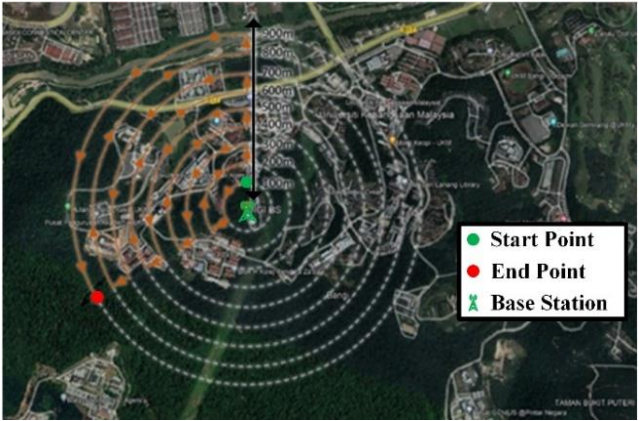

FIGURE 4. The measurement path.

This enabled comprehensive sampling of the propagation channel at diverse spatial points.

A dedicated repeat mission was conducted along a previously flown trajectory to ensure repeatability and data consistency. This provided a baseline for validating the collected data.

Essential signal quality metrics, including RSRP, RSRQ, SNR, RSSI, and Path Loss, were recorded during the measurements. Additionally, advanced network parameters were captured, including Channel State Information (CSI), Modulation and Coding Scheme (MCS), Physical Resource Blocks (PRBs) allocation, Block Error Rate (BLER), Hybrid Automatic Repeat Request (HARQ) retransmissions, and throughput at various protocol layers (physical, MAC, and RLC). These parameters were sampled at 2 Hz, comprehensively characterizing network behavior under aerial conditions.

This comprehensive measurement setup, combined with a structured flight strategy and controlled test conditions, ensured high-quality data collection for empirical A2G channel modeling. Additionally, the captured KPIs and network parameters provide a robust foundation for analyzing UAV cellular connectivity, enabling the development of an accurate and scalable propagation model tailored for aerial communications.

## IV. CHANNEL MODELING

To develop a robust predictive model that accurately characterizes LTE communication link performance for cellular-connected UAVs, we leveraged Keysight Nemo Handy to measure and model essential A2G propagation parameters, including RSRP, RSRQ, RSSI, and PL. As discussed in Sections I and III, these parameters are critical in assessing cellular link quality under aerial conditions.

Among these, RSRP is a fundamental metric for cell selection and handover decisions, as it quantifies the average received signal power per LTE Resource Block (RB). Meanwhile, RSSI represents the total received power across the entire occupied bandwidth, encompassing both desired and interfering signals. RSRQ, derived from RSRP and RSSI, accounts for the impact of cell load conditions and provides additional insights into network performance, particularly in handover scenarios [29].

$$RSRP[W] = \frac{1}{N}\sum_{n=1}^{N} P_n, \qquad (1)$$

where $N$ represents the count of received reference signals, and $P_n$ denotes the received power of the $n_{th}$ reference signal. However, relying solely on RSRP may not fully reflect actual signal quality since interference within the same frequency band is not considered explicitly.

To complement this limitation, RSRQ provides additional insights into signal quality conditions, accounting for interference effects from neighboring cells. It is computed based on RSSI and RSRP, as shown in (2):

$$RSRQ = N \times \frac{RSRP[W]}{RSSI[W]}. \qquad (2)$$

RSSI represents the power measured across the complete bandwidth of occupied RBs, encompassing intra-cell power, interference, and noise. It is mathematically defined as given in (3):

$$RSSI = \sum_{i=1}^{K} P_i + I + N_{th}, \qquad (3)$$

where, $P_i$ represents the power of the received reference signals over $K$ RBs, $I$ denotes the interference power from neighboring cells operating on the same frequency, and $N_{th}$ represents the thermal noise power within the channel bandwidth. RSRQ is a dimensionless quantity typically expressed in decibels (dB) and is often used to assess the quality of received signals in varying network conditions. A detailed signal quality classification based on measured RSRP and RSRQ values is presented in Table I, [30].

TABLE I. Signal status based on RSRP and RSRQ values.

| Signal Quality | RSRP | RSRQ | RSSI |
|---|---|---|---|
| Excellent | > -70 dBm | > -6 dB | > -65 dBm |
| Good | -70 ~ -80 dBm | -6 ~ -10 dB | -65 ~ -75 dBm |
| Medium | -80 ~ -90 dBm | -10 ~ -15 dB | -75 ~ -85 dBm |
| Weak | < -90 dBm | < -15 dB | < -85 dB |

## A. DATA PREPROCESSING AND FEATURE EXTRACTION

The data preprocessing and statistical analysis were initially conducted using NumPy [31] and Pandas [32] libraries for efficient data preprocessing and cleaning. The refined dataset was then utilized to train the proposed ML model using MATLAB R2023a. For additional details, readers are referred to [35] and [36]. Fig. 5 provides a detailed flowchart outlining the ML-based modeling approach for predicting RSRP, RSRQ, RSSI, and PL.

The initial dataset was structured as a matrix of dimensions 95,000 × 76, where each row represented a measurement sample containing key attributes such as the drone's location, timestamp, cellular network parameters, and received signal characteristics. The raw data underwent a multi-stage filtering process to extract only the samples associated with the serving BS while discarding unrelated or redundant records. Subsequently, data cleaning procedures were applied to eliminate outliers, missing values, and inconsistencies that could negatively impact model accuracy. Both the raw and cleaned datasets are accessible via [12].

To enhance predictive performance, four additional features were computed and integrated into the dataset as independent variables: 2D distance ($d_{2D}$), 3D distance ($d_{3D}$), azimuth angle ($\theta$), and elevation angle ($\beta$). The 2D propagation distance ($d_{2D}$) between the BS and UAV was derived using the Haversine formula, leveraging the GPS coordinates of both entities, as given in (4):

$$d_{2D} = 2 \times R \times arctan2(\sqrt{a}, \sqrt{1-a}), a = sin^2\left(\frac{lat_{BS} - lat_{UAV}}{2}\right) + cos(lat_{BS}) \times cos(lat_{UAV}) \times sin^2\left(\frac{lon_{BS} - lon_{UAV}}{2}\right), \quad (4)$$

where $R$ denotes the average radius of the Earth (6,371 km). $lat_{BS}$ and $lon_{BS}$ denote the latitude and longitude of the serving BS in decimal degrees, while $lat_{UAV}$ and $lon_{UAV}$ correspond to the latitude and longitude of the drone at each sampling point.

The 3D distance ($d_{3D}$) between the UAV and the serving BS accounts for both the horizontal (2D) separation and the vertical altitude difference. It is derived, as given in (5), using the Pythagorean theorem:

$$d_{3D} = \sqrt{d_{2D}^2 + (h_{UAV} - h_{BS})^2}, \quad (5)$$

where $h_{UAV}$ is the altitude of UAV and $h_{BS}$ is the height of the BS.

The elevation angle ($\beta$) was derived based on $h_{BS}$, $h_{UAV}$, and $d_{2D}$, as given in (6):

$$\beta = tan^{-1}\left(\frac{h_{UAV} - h_{BS}}{d_{2D}}\right) + \alpha, \quad (6)$$

where $\alpha$ represents the tilt angle of the BS's antenna, which was set to 4° in this study. Notably, since the UAV altitude always exceeded the BS antenna height in all measurement scenarios, the numerator in (6) remained positive, ensuring a consistent and meaningful elevation angle calculation.

The azimuth angle ($\theta$) represents the horizontal bearing from the BS to the UAV, measured clockwise from true north (0°). It is computed as per (7) using the inverse tangent (arctan2) function:

$$\theta = tan^{-1}\left(\frac{sin(\Delta\lambda).cos(lat_{UAV})}{cos(lat_{BS}).sin(lat_{UAV}) - sin(lat_{BS}).cos(lat_{UAV}).cos(\Delta\lambda)}\right), \quad (7)$$

where, $\Delta\lambda = lon_{UAV} - lon_{BS}$ is the longitude difference between the UAV and BS. Since azimuth angles are measured clockwise from true north, the computed value is normalized as given in (8):

$$\theta_{final} = (\theta + 360) \, mod \, 360. \quad (8)$$

This ensures that the azimuth angle remains within the valid range [0°, 360°].

Among these attributes, RSRP, RSRQ, RSSI, and PL were designated as dependent variables, forming the primary targets for prediction within the ML framework. To ensure feature consistency and improve model convergence, all computed attributes were normalized, effectively standardizing their scales and enhancing the learning efficiency of the ML model.

## B. MACHINE LEARNING MODELING FRAMEWORK

To comprehensively and accurately predict RSRP, RSRQ, RSSI, and PL, this study proposes a unique ML framework, that follows a triple-layer architecture, as illustrated in Fig. 5. This architecture comprises: (i) STW, (ii) EBT, and (iii) GPR aggregation. It combines the strengths of linear modeling simplicity (first layer), ensemble-based robustness against overfitting (second layer), and Gaussian processes' probabilistic inference capabilities (third layer).

The input to the model consists of a minimal yet highly impactful feature set—log-transformed 2-D and 3-D distances, azimuth angles, and elevation angles—ensuring computational efficiency and practical scalability. This compact feature design approach directly addresses the limitations of earlier single-layer models outlined in Section II. All modeling stages were trained and validated using a 10-fold cross-validation (CV) scheme to enhance robustness and mitigate overfitting risks.

Accordingly, for each modeling KPI (denoted as $Y \in \{PL, RSRP, RSRQ, RSSI\}$), the final prediction $Y$ is obtained as in (8) – (11):

$$\hat{Y} = GPR(X_{input}, \hat{Y}_{STW}, \hat{\varepsilon}_Y), \quad (8)$$

$$\hat{Y}_{STW} = STW(X_{input}), \quad (9)$$

$$\hat{\varepsilon}_Y = EBT(X_{input}), \quad (10)$$

where, $X_{input}$ represents the normalized input features, $\hat{Y}_{STW}$ is the linear prediction from the first layer, and $\hat{\varepsilon}_Y$ is the residual error prediction from the second layer.

In this modeling pipeline, the STW layer initially performs linear regression. During K-fold CV training stage, hyperparameter optimization of STW was performed to identify and retain significant linear relationships between input features and each of target KPIs. Next, residual errors, obtained by subtracting the actual normalized responses from initial STW layer predictions ($\varepsilon_Y = Y - \hat{Y}_{STW}$), were modeled in the second layer using EBT. This nonlinear modeling approach refined the initial predictions by capturing complex interactions and patterns not identified in the linear layer. Hyperparameter tuning was again performed via K-fold CV. The second layer's output reduces residual errors significantly.

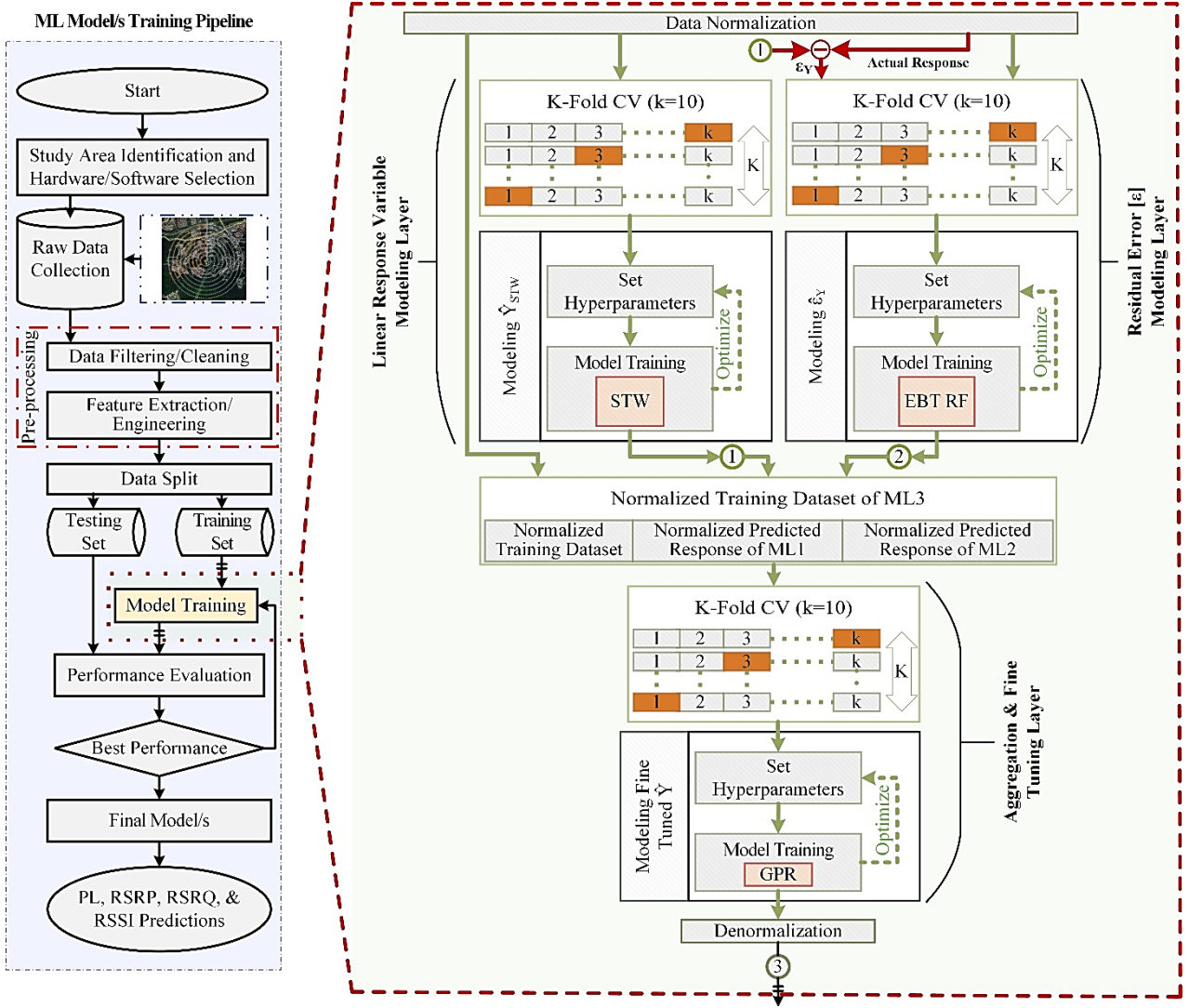

**FIGURE 5.** Flowchart of the proposed ML framework modeling pipeline.

The third and final modeling layer applies GPR to aggregate the STW and EBT outputs along with the original normalized inputs. Leveraging its probabilistic inference mechanism, GPR further fine-tunes the predictions and provides uncertainty estimates for each target. After training and optimization, predictions were denormalized to return to their original scale.

The complete modeling pipeline thus provides a rigorous methodological approach for accurately modeling and predicting cellular communication parameters within complex UAV operating scenarios, as will be proven in the next sections.

## V. PERFORMANCE EVALUATION

This section presents a comprehensive evaluation of the proposed triple-layer ML framework. As illustrated in Fig. 5, the dataset was split based on measurement location into 80% (8,910 samples) for training and 20% (2,227 samples) for testing to ensure robust performance analysis and minimize spatial bias. This design choice for data splitting guarantees that no overlapping or duplicate location data were used across the two subsets. Model training was performed exclusively on the training data, employing a 10-fold CV scheme to further enhance model robustness. Post-training, the framework was evaluated on the testing set to assess its generalization and predictive accuracy.

The evaluation covers the three modeling layers—STW, EBT, and GPR—across four performance KPIs: PL, RSRP, RSRQ, and RSSI. Model performance was assessed on both training and testing sets using six standard metrics: Mean Squared Error (MSE), Root Mean Squared Error (RMSE), Mean Absolute Error (MAE), Mean Arctangent Absolute Percentage Error (MAAPE), correlation coefficient (R), and coefficient of determination ($R^2$). These metrics collectively quantify prediction errors, residual variability, and linear association between predicted and actual values.

Specifically, MSE reflects the average squared deviation between actual and predicted values, while RMSE standardizes this error back to the original scale. MAE captures the absolute average of residuals and is generally known to be less sensitive to outliers, whereas MAAPE offers a robust alternative to MAPE, particularly for small or negative targets [33]. R quantifies the linear correlation strength between predictions and ground truth, while $R^2$ represents a robust measure of the prediction model accuracy while also showing how well the model's independent variables explain the variance of the dependent variable.

While lower MSE, RMSE, MAE, and MAAPE imply lower prediction error and higher accuracy, there is no specific scale for their values to decide on the quality of the prediction, i.e., how good or poor the model performs. Meanwhile, as indicated in [34], $R^2$ is the most informative metric in identifying the model prediction quality. In this regard, a negative $R^2$ value implies a poor prediction model, while a value from 0 to 1 implies that the prediction model is performing from worse to best. Due to its unity scale, $R^2$ is usually represented in percentile form to identify the model prediction accuracy rate.

For $n_S$ samples, these metrics are calculated as given in (9) to (14) [33], [35]–[38]:

$$MSE = \frac{1}{n_S}\sum_{i=1}^{n_S}(y_i - P_i)^2, \quad (11)$$

$$RMSE = \sqrt{MSE}, \quad (12)$$

$$MAE = \frac{1}{n_S}\sum_{i=1}^{n_S}|y_i - P_i|, \quad (13)$$

$$MAAPE = \left(\frac{1}{n_S}\sum_{i=1}^{n_S}sin^{-1}\left(\left|\frac{P_i - y_i}{P_i}\right|\right)\right) \times 100, \quad (14)$$

$$R = \frac{\sum_{i=1}^{n_S}(y_i - \bar{y})(P_i - \bar{P})}{\sqrt{\sum_{i=1}^{n_S}(y_i - \bar{y})^2 \sum_{i=1}^{n_S}(P_i - \bar{P})^2}}, \quad (15)$$

$$R^2 = 1 - \frac{\sum_{i=1}^{n_S}(y_i - P_i)^2}{\sum_{i=1}^{n_S}(y_i - \bar{y})^2}, \quad (16)$$

where $P_i$ is the predicted value for the $i$th sample, and $y_i$ is the corresponding actual measured value. Meanwhile, $\bar{y}$ and $\bar{P}$ are the mean of the actual and predicted datasets, respectively.

Table II reports the numerical performance on both training and testing sets, while Fig. 6 provides a side-by-side comparison of $R^2$ values across the three modeling phases. Additionally, visual insights including correlation plots, residual histograms, and predicted vs. measured response fit are drawn from Figs. 7 to 12, offering a qualitative understanding of the model behavior across the prediction pipeline.

## A. QUANTITATIVE AND VISUAL PERFORMANCE ANALYSIS ACROSS ALL MODELING LAYERS

As illustrated in Table II and Fig. 6, the predictive performance improves progressively from STW to EBT and finally to GPR. The initial STW layer serves as a simple linear estimator, establishing a foundational approximation. $R^2$ values in this stage range from 0.23 (RSRQ) to 0.44 (RSSI), i.e., 23% to 44%.

The corresponding measured vs. predicted plots in Fig. 7 and detailed performance assessment in Fig. 8 reveal a noticeable distribution from the ideal diagonal line. This is especially evident for RSRQ and PL, where the linear approximation struggles to capture complex spatial variability. The residual scatter plots and 2D histograms also exhibit wide error distributions, indicating poor concentration around zero and affirming the limitations of purely linear modeling.

TABLE II. Summary of model performance metrics across various modeling layers and datasets (training (TR) and testing (TS))

| KPI | Modeling Layer | Set | MSE | RMSE | MAE | MAAPE | R | $R^2$ |
|---|---|---|---|---|---|---|---|---|
| PL | STW | TR | 19.29 | 4.39 | 3.33 | 3.27 | 0.57 | 0.33 |
| | | TS | 17.97 | 4.24 | 3.23 | 3.13 | 0.56 | 0.31 |
| | EBT | TR | 0.99 | 0.99 | 0.66 | 29.77 | 0.98 | 0.95 |
| | | TS | 2.67 | 1.63 | 1.13 | 44.89 | 0.92 | 0.85 |
| | GPR | TR | 0.12 | 0.35 | 0.24 | 0.24 | 0.99 | 0.99 |
| | | TS | 2.54 | 1.59 | 1.04 | 1.02 | 0.95 | 0.90 |
| RSRP | STW | TR | 18.02 | 4.25 | 3.16 | 3.95 | 0.60 | 0.36 |
| | | TS | 16.57 | 4.07 | 3.04 | 3.76 | 0.61 | 0.37 |
| | EBT | TR | 0.76 | 0.87 | 0.52 | 25.46 | 0.98 | 0.96 |
| | | TS | 1.61 | 1.27 | 0.82 | 36.01 | 0.95 | 0.90 |
| | GPR | TR | 0.03 | 0.18 | 0.11 | 0.14 | 0.99 | 0.99 |
| | | TS | 1.25 | 1.12 | 0.69 | 0.86 | 0.98 | 0.95 |
| RSRQ | STW | TR | 8.69 | 2.95 | 2.11 | 15.19 | 0.48 | 0.23 |
| | | TS | 8.52 | 2.92 | 2.08 | 14.46 | 0.49 | 0.24 |
| | EBT | TR | 0.40 | 0.63 | 0.45 | 34.71 | 0.98 | 0.95 |
| | | TS | 1.11 | 1.06 | 0.78 | 48.81 | 0.93 | 0.87 |
| | GPR | TR | 0.07 | 0.27 | 0.20 | 1.50 | 0.99 | 0.99 |
| | | TS | 0.98 | 0.99 | 0.71 | 5.40 | 0.96 | 0.91 |
| RSSI | STW | TR | 5.35 | 2.31 | 1.70 | 3.81 | 0.66 | 0.44 |
| | | TS | 4.59 | 2.14 | 1.63 | 3.60 | 0.66 | 0.43 |
| | EBT | TR | 0.48 | 0.69 | 0.46 | 38.69 | 0.96 | 0.91 |
| | | TS | 0.86 | 0.93 | 0.70 | 51.80 | 0.90 | 0.81 |
| | GPR | TR | 0.07 | 0.27 | 0.19 | 0.41 | 0.99 | 0.99 |
| | | TS | 0.75 | 0.87 | 0.65 | 1.42 | 0.95 | 0.91 |

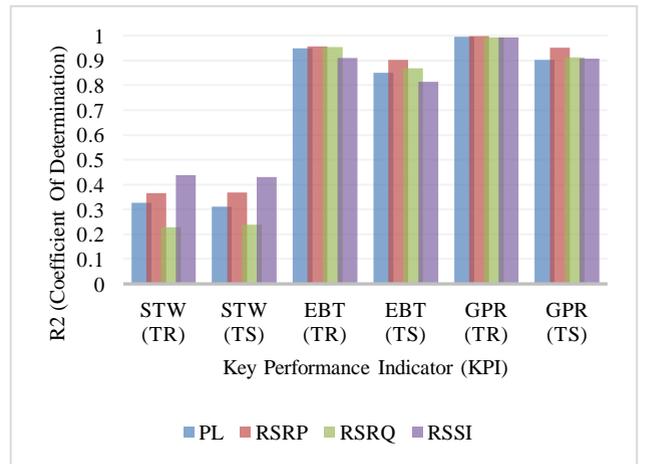

FIGURE 6. $R^2$ comparison across all modeling layers (STW, EBT, GPR) for PL, RSRP, RSRQ, and RSSI considering both training (TR) and testing (TS) sets.

The second layer (EBT) substantially elevates performance by modeling the nonlinear structure of residuals. For instance, the $R^2$ for PL increases from 33% in STW to 95% in EBT (training set), with similar gains observed across the other KPIs. Fig. 9 illustrates the CDF plots of the prediction residual ($\varepsilon_Y$ vs $\hat{\varepsilon}_Y$), while Fig. 10 provides correlation and residual assessment.

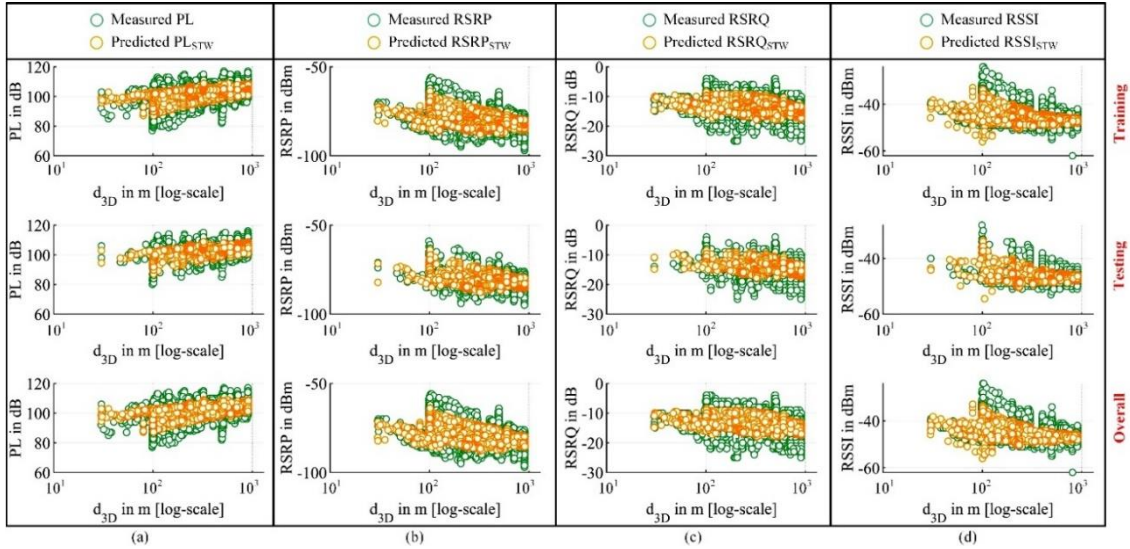

**FIGURE 7.** Measured ($Y$) versus predicted ($\hat{Y}_{STW}$) KPIs from the STW modelling layer, specifically: (a) PL, (b) RSRP, (c) RSRQ, and (d) RSSI, across the training, testing, and overall dataset.

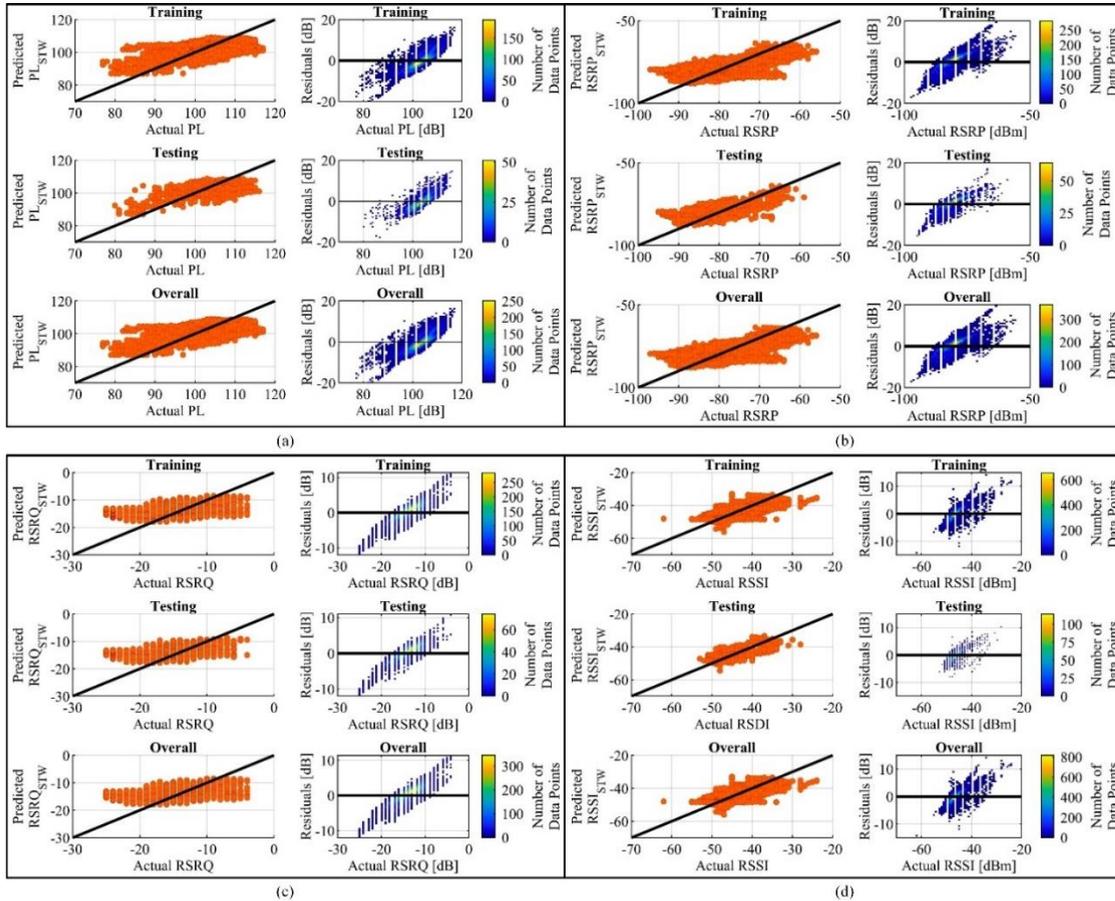

**FIGURE 8.** Performance of STW model across training, testing, and overall datasets for (a) PL, (b) RSRP, (c) RSRQ, and (d) RSSI. In all figures, the left side presents actual versus predicted response variable correlation scatter plots, while the right side displays residuals versus the actual response variable scatter plots. The latter also shows a 2D histogram that illustrates the density of residuals spread across the range of the actual response variable.

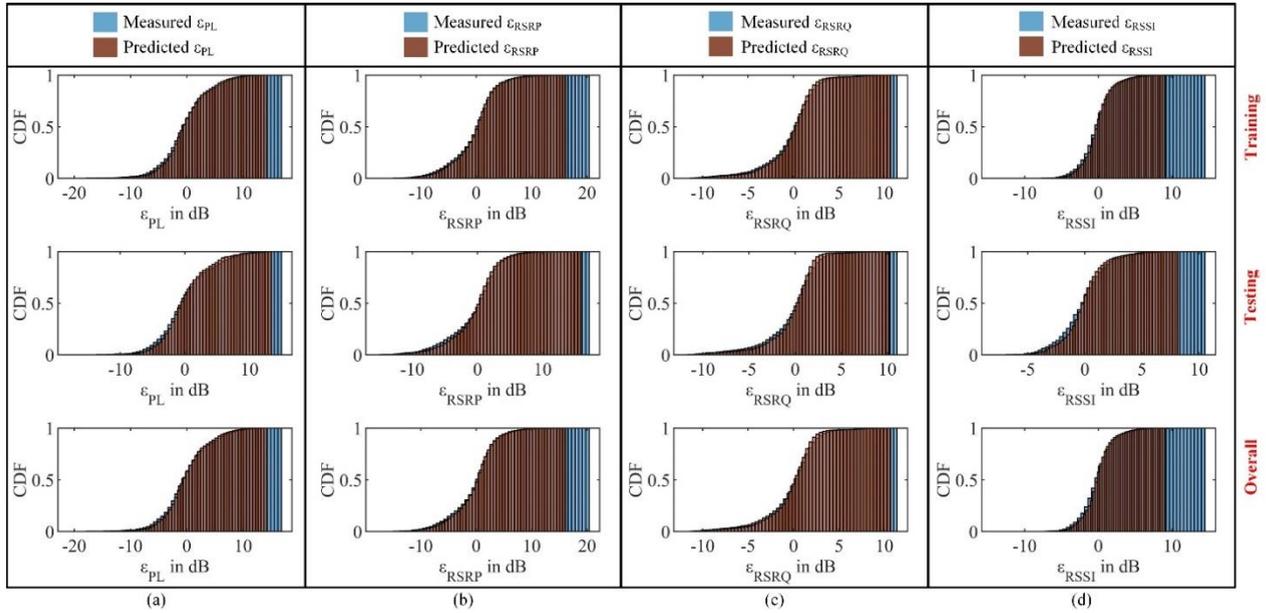

FIGURE 9. CDF of measured ($\varepsilon_Y$) versus predicted ($\hat{\varepsilon}_Y$) KPIs from the EBT modelling layer, specifically: (a) PL, (b) RSRP, (c) RSRQ, and (d) RSSI, across the training, testing, and overall dataset.

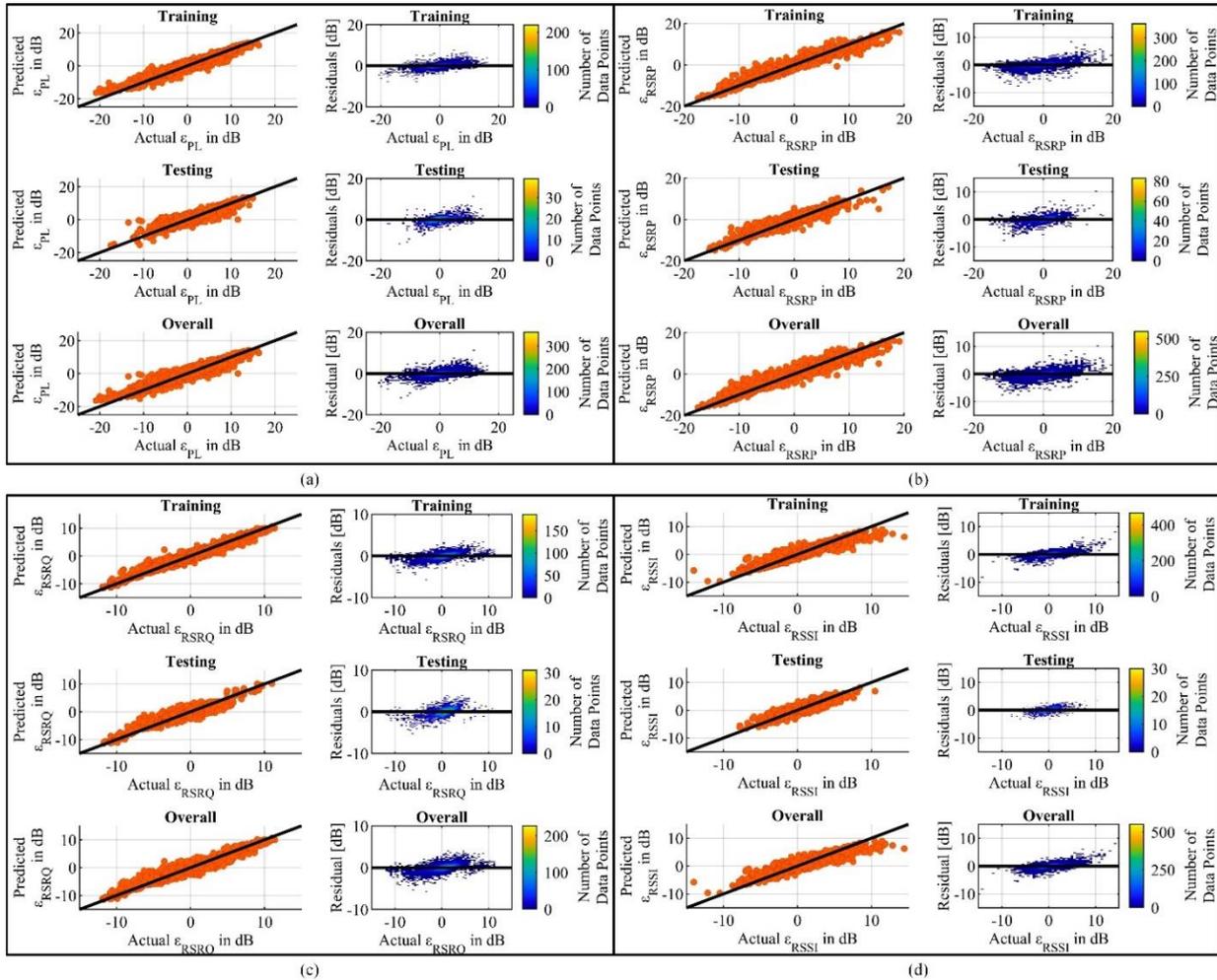

FIGURE 10. Performance of EBT model across training, testing, and overall datasets for (a) PL, (b) RSRP, (c) RSRQ, and (d) RSSI. In all figures, the left side presents actual versus predicted response variable correlation scatter plots, while the right side displays residuals versus the actual response variable scatter plots. The latter also shows a 2D histogram that illustrates the density of residuals spread across the range of the actual response variable.

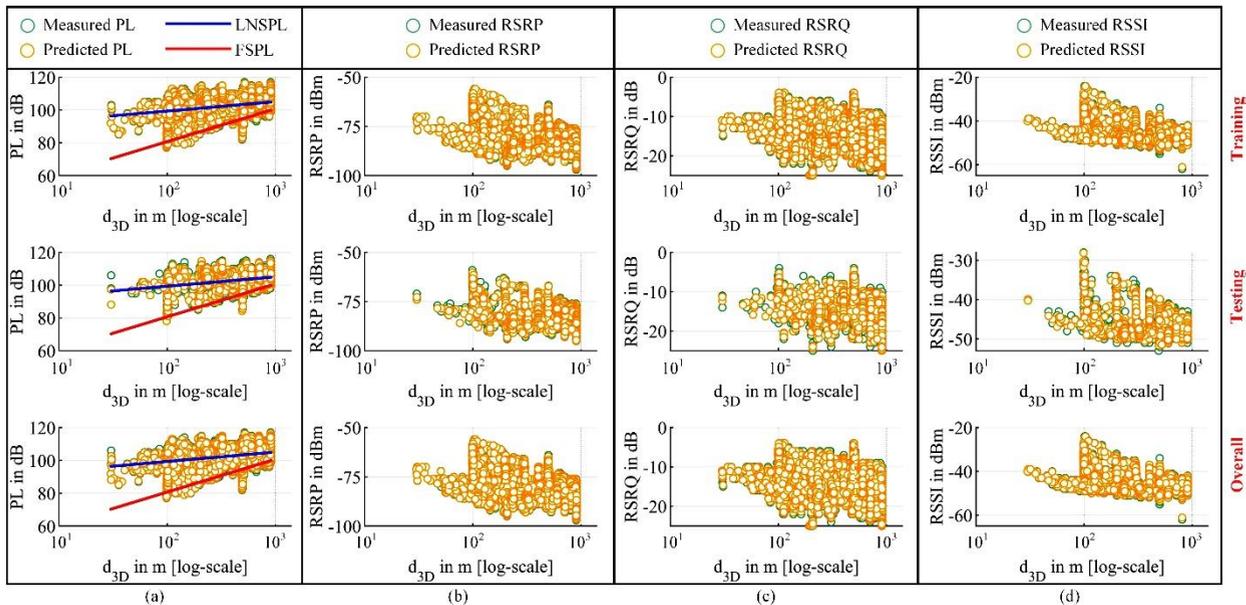

**FIGURE 11.** Measured ($Y$) versus predicted ($\hat{Y}$) KPIs from the GPR modelling (aggregation) layer, specifically: (a) PL, (b) RSRP, (c) RSRQ, and (d) RSSI, across the training, testing, and overall dataset.

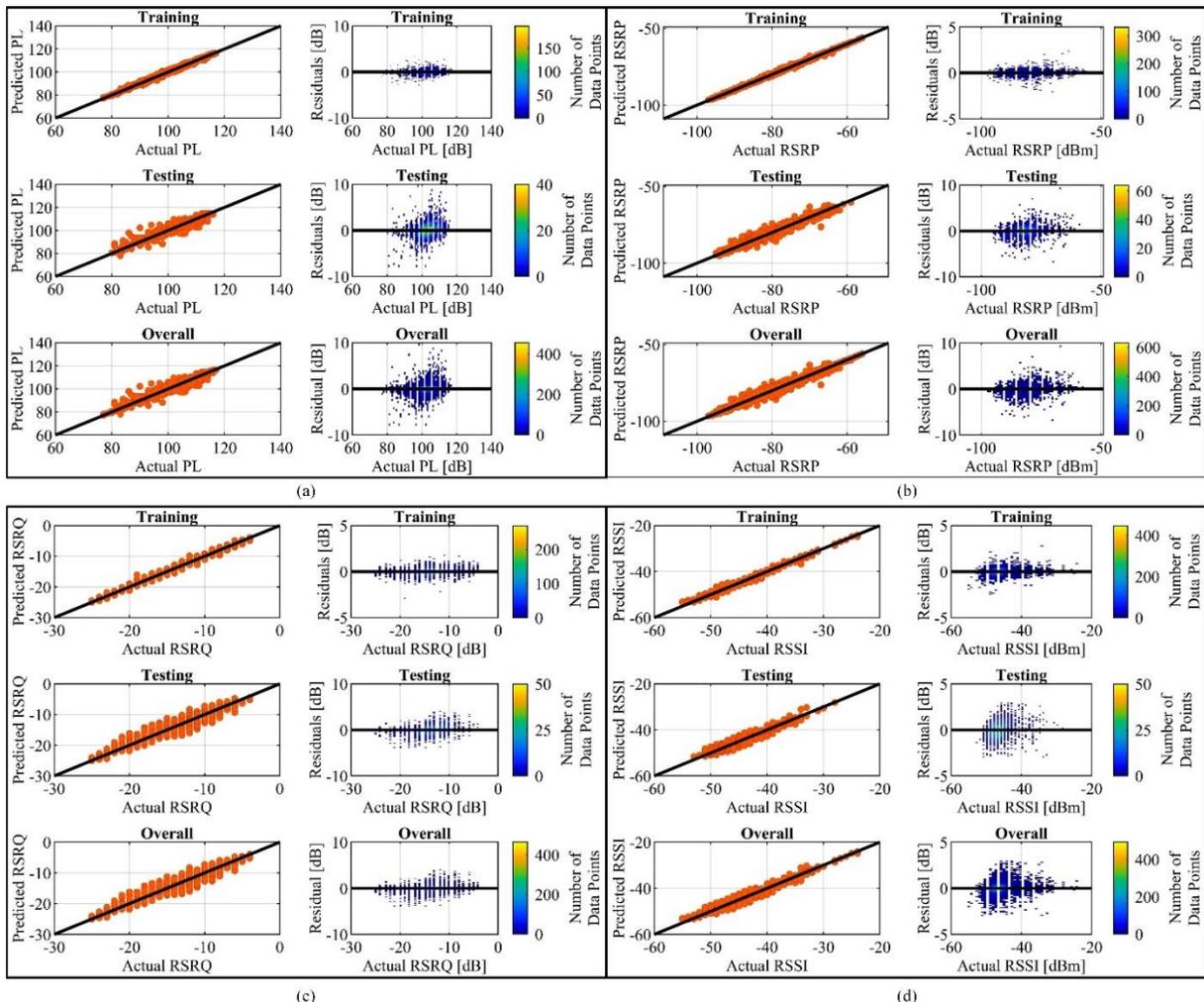

**FIGURE 12.** Performance of GPR model across training, testing, and overall datasets for (a) PL, (b) RSRP, (c) RSRQ, and (d) RSSI. In all figures, the left side presents actual versus predicted response variable correlation scatter plots, while the right side displays residuals versus the actual response variable scatter plots. The latter also shows a 2D histogram that illustrates the density of residuals spread across the range of the actual response variable.

The residuals are more tightly clustered around zero, and the histogram densities become more peaked, especially for PL and RSRP, reflecting improved model stability and accuracy. However, some variability remains for RSRQ and RSSI, owing to their susceptibility to noise and cross-cell interference.

The final GPR aggregation layer yields further refinement by probabilistically combining the outputs from STW and EBT. Table II and Fig. 6 indicate that R² reaches around 99% for all KPIs in the training set and remains above 90% in testing. The measured vs. predicted plots in Fig. 11, together with the performance analysis of correlation and residual histogram in Fig. 12, demonstrate near-perfect alignment with ground truth. Residuals are densely concentrated around zero, and histogram peaks become narrower, ranging in most cases between ±10 dB, particularly for RSRP. These improvements highlight the GPR layer's ability to correct remaining bias and capture subtle spatial nuances. Similar applies to other KPIs while posing a slight challenge due to its compounded formulation, yet the GPR layer still pushes R² for all KPIs beyond 90% in the testing set with visibly reduced residual variance and stronger prediction fits. It should also be noted that for reference, the PL plots shown in both Fig. 11 were compared against reference PL models, namely free space PL (FSPL) and log-normal shadowing PL (LNSPL) models. In both cases, it can be seen that the final model yields the best fit against the measured PL samples, therefore further affirming the superior performance of the proposed models.

### B. OBSERVATIONS ON GENERALIZATION AND TESTING PERFORMANCE

An important aspect of the model evaluation lies in assessing how well the trained models generalize to new data. While the proposed triple-layer ML framework demonstrates high predictive performance across all layers, a careful assessment of the testing results reveals slight but expected performance drops compared to the training set.

This behavior is most apparent in the EBT modeling layer. For instance, MAAPE for PL increases from 29.77% in the training set to 44.89% in the testing set, and similar trends are observed for the remaining KPIs. These discrepancies are attributed to the inherent spatial diversity and signal variability in UAV-based propagation environments, where factors such as obstructions, terrain irregularities, and multipath conditions can differ substantially across measurement locations. Given that the data splitting was based on unique locations, with no overlap between training and testing locations, such differences are expected.

Nevertheless, the observed degradation remains within acceptable bounds. Even at its lowest, the testing R² for RSRQ, which is considered the most challenging KPI due to its compounded nature, still exceeds 91%. The remaining KPIs, including RSRP, PL, and RSSI, maintain testing R² values above 90%, affirming the model's ability to retain generalization performance despite changing conditions. These values indicate that over 90% of the variance in the actual KPI values is accurately captured by the model predictions on a new dataset.

The visual results in Figs. 7 to 12 corroborate this analysis. While the testing scatter plots exhibit slightly broader distributions than the training plots, they remain tightly centered along the ideal prediction line. Residual distributions also imply significant performance, mostly concentrated around zero with minimal skewness, particularly in the GPR stage. The consistency in the residual spread and the strong predictive alignment confirm that the framework is not overfitting but rather exhibiting appropriate generalization.

In summary, although minor reductions in prediction accuracy are observed across the testing set, they reflect the realistic variability of UAV-based communication environments and remain within acceptable performance thresholds. The consistently high R² values, exceeding 90% across all KPIs, combined with low residual dispersion, highlight the proposed framework's robustness and strong predictive reliability under unseen spatial conditions.

## VI. MODEL COMPLEXITY AND USER-ORIENTED CONFIGURATION STRATEGY

A practical ML-based framework is expected to deliver high predictive accuracy and exhibit computational efficiency conducive to real-time deployment. Accordingly, this section discusses the prediction complexity of the proposed triple-layer ML framework and compares it against other typical modeling approaches, with a focus on prediction speed and resource demand. The discussion further outlines configuration options tailored to end users' operational constraints and priorities.

Compared with our previous work in [38] that proposes a dual-layer model stacking a FSPL baseline with STW/SVM followed by an EBT or ANN model to predict shadowing impact and relies on 14 input features, with ray-tracing among the considered feature engineering steps, to predict PL only, the present triple-layer pipeline trains on only four geometric features and simultaneously predicts PL, RSRP, RSRQ, and RSSI. This approach cuts feature dimensionality by more than 70 %, removes categorical-encoding overhead, and underpins the latency- and accuracy-tuned configurations detailed below.

### A. PREDICTION SPEED AND MODEL OPTIMIZATION

Upon conducting final simulation profiling on the complete triple-layer modeling pipeline, it was observed that the default configuration—particularly the second-layer EBT model—introduces non-negligible latency. Specifically, the prediction speed under default settings (i.e., minimum leaf size of 1, number of variables to sample, and number of learners which was optimized to 375 for maximum prediction accuracy) was measured at approximately 5,325 observations per second (obs/sec) across the modeled KPIs (PL, RSRP, RSRQ, and RSSI). While this performance is sufficient for most offline analytical tasks, it may not meet the responsive requirements of latency-sensitive applications, such as real-time monitoring or autonomous handover decisions.

To address this, a lightweight variant was introduced by tuning the EBT model hyperparameters, notably reducing the number of learners to 60 and capping the minimum leaf size to 8 while considering all variables for sampling and maximum number of tree splits ranging from 2,378 up to 6,540. This adjustment yielded a significant complexity reduction, increasing the average prediction speed to approximately 19,476 obs/sec, while only incurring a slight degradation in prediction accuracy—namely, an average decrease of ≈1–2 % in $R^2$ and an RMSE increase of < 0.2 dB across all KPIs for testing set. Given the inherent variability and measurement noise within the UAV-based cellular channel, these deviations remain within acceptable performance margins.

### B. COMPARATIVE EVALUATION AGAINST TYPICAL MODELING PIPELINES

While most ML-based wireless propagation studies emphasize accuracy metrics such as RMSE or $R^2$, only a few explicitly report prediction speed or computational load. In this regard, several simulations were done considering typical modeling pipelines, having a single modeling layer with various model types that serve as a relevant benchmark to compare the proposed models against. In these simulations, input features were directly fed into various ML models, including LR, DT, SVM, Ensemble, GPR, and ANN models with various hyperparameter configurations. Accordingly, a total of 16 models were trained and tested using the earlier prepared training and testing datasets. These models were then compared against this work's proposed ML-based models in terms of model accuracy and prediction speed to highlight model complexity.

In addition, the study in [39], offer prediction speed comparisons across multiple ML models for signal coverage mapping which were considered for further comparison. Their findings show that SVM and GPR-based models respectively achieve 11,000 obs/sec and 8,600 obs/sec, with RMSE values of 3.98 dB and 1.97 dB for RSRP modeling. Although their reported accuracies are commendable, their models are designed for single KPI prediction under more constrained environments. In contrast, the proposed framework achieves multi-KPI prediction (four KPIs concurrently) with superior accuracy (testing RMSE as low as 0.87 dB and $R^2$ as low as 90%) and competitive speed, particularly under the optimized configuration. These results confirm that the triple-layer architecture strikes a favorable balance between complexity and accuracy, benefiting from the modularity and interpretability of its components. This is further supported by the simulations done for other considered models, as summarized in Table III. It can be clearly noted that the proposed models offer superior accuracy and prediction speed against all compared models. While some models showed higher prediction speeds, they lack prediction accuracy, which is way below that of the proposed approach, for both variants (accuracy or complexity optimized variants). Hence, the proposed models serve as balanced models that may be used depending on the targeted application requirements, as highlighted in the next subsection. The results underscore the flexibility of the proposed framework in accommodating varying accuracy-speed trade-offs while outperforming conventional ML models in both aspects.

### C. PRACTICAL DEPLOYMENT RECOMMENDATIONS

From a deployment perspective, two configurations are proposed to accommodate the diverse operational contexts of prospective users:

1. *Latency-optimized configurations*: Suitable for applications requiring fast, near real-time predictions (e.g., UAV-assisted field scanning, adaptive coverage mapping). This version achieves up to 19,476 obs/sec while retaining acceptable accuracy (RMSE: 0.97 – 1.8 dB; $R^2$: 88% – 94%), making it ideal for edge-computing scenarios.
2. *Accuracy-prioritized configurations*: Recommended for offline analysis or cloud-hosted systems where computational resources are less constrained. This version retains full EBT complexity, offering peak accuracy (RMSE: 0.87 – 1.59 dB; $R^2$: 90% – 95%) at 5,325 obs/sec.

While in the current study, the latter choices are made manual, in future works, the model structure can be modified to inherently support dynamic switching between these configurations based on application context. By maintaining identical STW and GPR layers, a runtime control policy may be implemented to toggle EBT complexity in response to system load or latency thresholds.

TABLE III. Comparative Performance of ML-based models (using conventional single layer approach) and current work in terms of $R^2$, RMSE, and prediction speed (organized as minimum to maximum achievable ranges)

| Model Number | Model Type \| Configuration | RMSE (TR) | $R^2$ (TR) | RMSE (TS) | R2 (TS) | Prediction Speed (obs/sec) |
|---|---|---|---|---|---|---|
| 1 | LR \| Linear | 2.6 - 4.6 | 0.14 - 0.27 | 2.4 - 4.4 | 0.15 - 0.27 | 52,923 – 148,501 |
| 2 | LR \| Interactions | 2.5 - 4.4 | 0.23 - 0.37 | 2.3 - 4.2 | 0.24 - 0.37 | 59,802 – 118,534 |
| 3 | LR \| Robust | 3.1 - 8.5 | -1.56 - -0.03 | 2.8 - 7.7 | -1.25 - 0.04 | 32,206 – 169,884 |
| 4 | LR \| STW | 2.5 - 4.4 | 0.23 - 0.37 | 2.3 - 4.2 | 0.24 - 0.37 | 66,896 – 83,613 |
| 5 | DT \| Coarse Tree | 1.9 - 2.9 | 0.64 - 0.75 | 2.5 - 2.9 | 0.64 - 0.68 | 68,473 – 279,817 |
| 6 | SVM \| Linear | 2.7 - 4.7 | 0.11 - 0.26 | 2.5 - 4.5 | 0.12 - 0.25 | 16,891 – 32,922 |
| 7 | SVM \| Quadratic | 2.4 - 4.3 | 0.26 - 0.38 | 2.2 - 4 | 0.28 - 0.42 | 20,058 – 30,046 |
| 8 | SVM \| Cubic | 3 - 5.2 | -1.29 - 0.69 | 2.6 - 4.1 | 0.12 - 0.69 | 6,902 – 87,518 |
| 9 | SVM \| Medium Gaussian | 2.1 - 3.7 | 0.51 - 0.63 | 2 - 3.6 | 0.49 - 0.63 | 5,817 – 9,987 |
| 10 | SVM \| Coarse Gaussian | 2.5 - 4.2 | 0.26 - 0.4 | 2.3 - 4.1 | 0.27 - 0.41 | 5,298 – 7,492 |
| 11 | Ensemble \| Boosted Trees | 2.2 - 5.5 | -0.04 - 0.59 | 2.2 - 5.5 | -0.14 - 0.58 | 34,707 – 56,046 |
| 12 | Ensemble \| Bagged Trees | 1.7 - 2.7 | 0.69 - 0.78 | 1.5 - 2.4 | 0.73 - 0.77 | 7,216 – 28,879 |
| 13 | GPR \| Squared Exponential | 1.8 - 3 | 0.64 - 0.72 | 1.7 - 3 | 0.64 - 0.72 | 3,440 – 4,633 |
| 14 | GPR \| Matern 5/2 | 1.8 - 2.9 | 0.66 - 0.74 | 1.6 - 2.8 | 0.67 - 0.75 | 2,779 – 3,206 |
| 15 | GPR \| Rational Quadratic | 1.8 - 2.8 | 0.67 - 0.76 | 1.6 - 2.6 | 0.69 - 0.75 | 2,256 – 3,770 |
| 16 | ANN | 1.9 - 3.5 | 0.57 - 0.68 | 1.6 - 3.2 | 0.62 - 0.72 | 80,881 – 262,185 |
| This Work | Triple layer (Latency-optimized configurations) | 0.81 – 1.32 | 0.94 – 0.98 | 0.97 – 1.8 | 0.88 – 0.94 | 19,476 |
| | Triple layer (Accuracy-prioritized configurations) | 0.18 – 0.35 | 0.99 | 0.87 – 1.59 | 0.90 – 0.95 | 5,325 |

## VII. CONCLUSION

This study introduced a structured triple-layer ML framework designed to accurately predict PL, RSRP, RSRQ, and RSSI in real-world UAV-assisted measurement environments. The proposed approach combines STW, EBT, and GPR in a sequential modeling pipeline, where each layer contributes distinct enhancements—from capturing linear trends to refining nonlinear residuals and finally performing probabilistic aggregation.

The model leverages a minimal yet effective input feature set and was trained using a 10-fold CV, with data split spatially to ensure realistic generalization. Performance was evaluated using six standard metrics across training and testing sets. Results showed a clear progression in accuracy across the three modeling phases, with GPR achieving final testing $R^2$ values above 90% for all KPIs and MAAPE values remaining consistently below 1.5%.

Visual assessments reinforced the numerical findings, with measured vs. predicted plots and residual histograms indicating high alignment, low error variance, and minimal prediction bias. While minor drops were noted in testing performance, they remained within acceptable margins and are attributable to the non-stationary nature of UAV measurement environments.

Overall, the proposed ML framework demonstrates strong generalization, high prediction accuracy, and effective scalability. It also offers flexible deployment options by providing both accuracy-prioritized and latency-optimized configurations, allowing adaptation to diverse application requirements. It serves as a practical and robust solution for channel modeling in UAV-based wireless systems. Future work can consider incorporating environmental or temporal variables, expanding to additional KPIs, or developing adaptive mechanisms to switch between complexity levels based on real-time processing constraints dynamically.

## ACKNOWLEDGMENT

This work was supported by Collaborative Research in Engineering, Science and Technology (CREST) and Aerodyne Group under grant reference number T23C2-19. We also acknowledged various forms of contributions from Universiti Kebangsaan Malaysia and Sunway University in getting this research work being published.

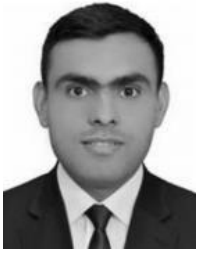

**HAIDER A.H. ALOBAIDY** (Member, IEEE) is a lecturer at the College of Information Engineering, Al-Nahrain University, Iraq. Prior to his current role, he served as a postdoctoral researcher at Universiti Kebangsaan Malaysia (UKM), where he further honed his expertise in wireless communication systems and related fields. He received a M.Sc. degree from Al-Mustansiriyah University, Iraq (2016) and a Ph.D. degree in Electrical, Electronics, and System Engineering from UKM (2022). His research interests include wireless communication, wireless sensor networks, IoT, machine learning, and channel propagation modeling and estimation.

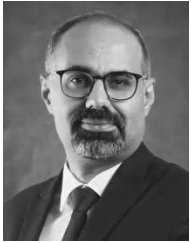

**MEHRAN BEHJATI** is currently a lecturer at Sunway University. Previously, he served as a postdoctoral researcher at the National University of Malaysia (UKM) and worked as a research associate at Iran's National Elites Foundation. Mehran obtained a B.Eng. in Electrical and Electronic Engineering from Azad University of Iran in 2009, followed by M.Eng. and Ph.D. degrees in Communication and Computer Engineering from UKM in 2013 and 2017, respectively. His research interests encompass wireless networks, IoT, edge intelligence, applied machine learning, and technology-driven solutions for environmental conservation.

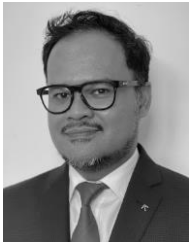

**ROSDIADEE NORDIN** (Senior Member, IEEE) received the B.Eng. degree from the Universiti Kebangsaan Malaysia, in 2001, and the Ph.D. degree from the University of Bristol, U.K., in 2011. He is currently a Professor at the Department of Engineering, Faculty of Engineering and Technology, Sunway University. His research interest includes Beyond 5G wireless communications, specifically focusing on advanced wireless transmission techniques and channel modeling, aerial wireless communications, and wireless communications for the Internet of Things applications. He is the recipient of the Leadership in Innovation Fellowship, a technopreneur program under the Royal Academy of Engineering, United Kingdom, in 2021, and the recipient of Top Research Scientists Malaysia (TRSM), a prestigious award under the Academy of Science Malaysia, in 2020.

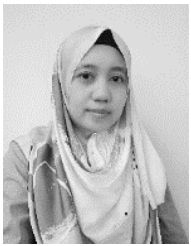

**NOR FADZILAH ABDULLAH** (Member, IEEE) is an Associate Professor of Electrical, Electronic & Systems Engineering at UKM. She received a M.Sc. degree from the University of Manchester (2003) and a Ph.D. degree in Electrical and Electronic Engineering from the University of Bristol (2012). Her research interests include 5G networks, vehicular networks, space-time coding, fountain code, and channel propagation modeling and estimation.